\documentclass[aps,prd,twocolumn,groupedaddress,nofootinbib]{revtex4}  % for review and submission

\usepackage{graphicx}  % needed for figures
\usepackage{dcolumn}   % needed for some tables
\usepackage{bm}        % for math
\usepackage{amssymb,amsfonts,amsmath,physics}   
\usepackage{cancel}
\usepackage{mathtools}
\usepackage{slashed}
\usepackage{url}
\usepackage{tikz-feynman}
\tikzfeynmanset{compat=1.1.0}
\usetikzlibrary{arrows}
\tikzset{	
	vertex/.style={circle,draw, minimum size=1.5em},	
	edge/.style={->,> = latex'}	
}
\usepackage[bookmarks, breaklinks, colorlinks,urlcolor=blue, citecolor=red, 
linkcolor=blue]{hyperref}
\usepackage[normalem]{ulem}

\newcommand{\be}{\begin{eqnarray*}}
	\newcommand{\ee}{\end{eqnarray*}}

\newcommand{\bee}{\begin{eqnarray}}
	\newcommand{\eee}{\end{eqnarray}}
\newcommand{\beeq}{\begin{equation}}
	\newcommand{\eeq}{\end{equation}}

\renewcommand{\vec}{\bf}

\usepackage{xspace}

\newcommand{\ba}{\begin{array}}
	\newcommand{\ea}{\end{array}}
\newcommand{\bd}{\begin{displaymath}}
	\newcommand{\ed}{\end{displaymath}}
\newcommand{\besub}{\begin{subequations}}
	\newcommand{\eesub}{\end{subequations}}
\newcommand{\bea}{\begin{eqnarray}}
	\newcommand{\eea}{\end{eqnarray}}

\def\q2 {q^2}

\usepackage{tikz}
\usetikzlibrary{arrows,shapes}
\usetikzlibrary{trees}
\usetikzlibrary{matrix,arrows} 				% For commutative diagram
\usetikzlibrary{positioning}				% For "above of=" commands
\usetikzlibrary{calc,through}				% For coordinates
\usetikzlibrary{decorations.pathreplacing}  % For curly braces
\usepackage{pgffor}							% For repeating patterns

\usetikzlibrary{decorations.pathmorphing}	% For Feynman Diagrams
\usetikzlibrary{decorations.markings}
\tikzset{
	vector/.style={decorate, decoration={snake}, draw},
	provector/.style={decorate, decoration={snake,amplitude=2.5pt}, draw},
	antivector/.style={decorate, decoration={snake,amplitude=-2.5pt}, draw},
	fermion/.style={draw=black, postaction={decorate},
		decoration={markings,mark=at position .55 with {\arrow[draw=black]{>}}}},
	fermionbar/.style={draw=black, postaction={decorate},
		decoration={markings,mark=at position .55 with {\arrow[draw=black]{<}}}},
	fermionnoarrow/.style={draw=black},
	gluon/.style={decorate, draw=black,
		decoration={coil,amplitude=4pt, segment length=5pt}},
	scalar/.style={dashed,draw=black, postaction={decorate},
		decoration={markings,mark=at position .55 with {\arrow[draw=black]{>}}}},
	scalarbar/.style={dashed,draw=black, postaction={decorate},
		decoration={markings,mark=at position .55 with {\arrow[draw=black]{<}}}},
	scalarnoarrow/.style={dashed,draw=black},
	electron/.style={draw=black, postaction={decorate},
		decoration={markings,mark=at position .55 with {\arrow[draw=black]{>}}}},
	bigvector/.style={decorate, decoration={snake,amplitude=4pt}, draw},
}

\tikzstyle{block} = [draw, rectangle, 
minimum height=3em, minimum width=6em]

\begin{document}
%\preprint{IP/BBSR/2015-4}
\title{Effects of Electroweak Symmetry Breaking on Axion Like Particles as Dark Matter}

\author{Soumen Kumar Manna}
\email{skmanna2021@gmail.com}
\affiliation{Department of Physics, Indian Institute of Technology Guwahati, Assam-781039, India}

\author{Arunansu Sil}
\email{asil@iitg.ac.in}
\affiliation{Department of Physics, Indian Institute of Technology Guwahati, Assam-781039, India}

\begin{abstract} 

Axion like particles (ALPs), the pseudo Nambu-Goldstone bosons associated to the spontaneous breaking 
of global symmetry, have emerged as promising dark matter candidates. Conventionally, in the context of misalignment mechanism, the non-thermally produced ALPs happen to stay frozen due to Hubble friction initially and at a later stage, they begin to oscillate (before matter-radiation equality) at characteristic frequencies defined by their masses and behaving like cold dark matter. In this work, we study the influence of electroweak symmetry breaking (EWSB), through a higher order Higgs portal interaction, on the evolution of ALPs. Such an interaction is found to contribute partially to the ALP's mass during EWSB, thereby modifying oscillation frequencies during EWSB as well as impacting upon the existing correlation between the scale of symmetry breaking and their masses. The novelty of the work lies in broadening the relic satisfied parameter space so as to probe it in near future via a wide range of experiments. 

\end{abstract}

\maketitle
\section{Introduction}
Axion like particles (ALPs) are generically classified as the pseudo Nambu-Goldstone bosons (pNGB) associated to the spontaneous breaking of a global symmetry, prevalent in many extensions of the Standard Model (SM) $e.g.$ in superstrings theories \cite{Svrcek:2006yi,Arvanitaki:2009fg,Demirtas:2018akl}. Though they are analogous to the pNGB of the $U(1)$ Peccei Quinn symmetry, originally introduced to solve the strong CP problem of QCD \cite{Peccei:1977hh,Peccei:1977ur,Weinberg:1977ma,Wilczek:1977pj}, ALPs belong to a more general class and hence not necessarily be connected to the solution of the strong CP problem. They are typically very light, neutral pseudo-scalar particles having at most a derivative coupling with the Standard Model at an effective level, suppressed by the scale of spontaneous symmetry breaking ($f_a$) of the global symmetry (also referred to as the ALP decay constant), thereby emerging as ideal candidates for explaining the dark matter of the Universe. The mass of an ALP may originate from non-perturbative instanton effect in a strongly interacting hidden sector (analogous to the QCD axion) as well as from explicit symmetry breaking. Correspondingly, contrary to the QCD axion, an ALP does not in general carry any specific relation between its mass and the decay constant. As a result, their mass and decay constant may span over a wide range making them attractive from detection point of view. 

Most commonly in the literature, an ALP is considered to be produced non-thermally via the so called $\it{misalignment}$ mechanism \cite{Preskill:1982cy,Abbott:1982af,Dine:1982ah,Turner:1983he,Arias:2012az}. The ALP, considered as a classical field in this context, is expected to have an initial non-zero field value and gets frozen there till the Hubble ($\mathcal{H}$) induced friction remains larger that its mass ($m_a$). Once $\mathcal{H} \sim m_a$, the ALP starts to oscillate at a temperature $T_{{osc}}$ about the minimum of a periodic potential characteristics of a pNGB. Such an oscillatory field mimics as cold dark matter as the associated energy density ($\rho_a$) scales as $R^{-3}$, hence behaving like ordinary matter ($R$ is the scale factor of the Universe) provided it remains stable over cosmological time scale. 
The relic density satisfaction of such ALP provides a standard correlation involving $m_a$ and $f_a$ which turns out to be well restricted by several cosmological constraints as well from ALP search experiments as summarised in \cite{Arias:2012az,Cadamuro:2011fd,Ringwald:2012hr,ParticleDataGroup:2022pth,Marsh:2017hbv}. Considering all such aspects, it turns out that the ALPs being DM should be very light (below keV). 
%[arXiv:220513549/1110.2895].

In this work, we propose to include a higher order shift symmetry breaking Higgs portal interaction of ALPs which contributes to its mass ($m_{aE}$) after the electroweak symmetry breaking (EWSB) on top of its existing mass, $m_{a0}$, presumably followed from non-perturbative instanton effect. Inclusion of such additional mass $m_{aE}$ is expected to modify the frequency of ALP oscillation after the EWSB. This would have profound effect on the relic density and hence on the ALP parameter space in $m_a, f_a$ plane, where $m_a^2 = {m_{a0}}^2 + {m_{aE}}^2$. We find that depending upon the onset of conventional ALP oscillation (connected with its mass $m_{a0}$ only) before or after the EWSB and associated modification of it after EWSB, the standard correlation between ALP mass and decay constant can be altered significantly leading to opening up of otherwise excluded (restricted) parameter space ($e.g.$ keV-GeV range) for ALPs. 

There are some studies focusing on modification of the PQ axion or ALP potential. For example, in ref. \cite{Jeong:2022kdr}, the authors examine the effect of tiny (limited by neutron electric dipole moment) explicit Peccei-Quinn (PQ) symmetry breaking on the PQ axion dynamics and its role as DM. In this case, the PQ axion initially starts oscillating about a wrong minimum guided by the explicit PQ breaking term and afterward it oscillates about the true minimum leading to a modification of the conventional PQ parameter space. Ref. \cite{Li:2023xkn} recently analyses the effect of introducing a PQ breaking term on axiverse that encompasses the PQ axion, an ALP and a hypothetical mixing between them. 
In another work, ref. \cite{Chatrchyan:2023cmz} discusses the effect of adding a non-periodic potential to ALP and thereby find the possibility of accommodating a large misalignment angles which may change the conventional $(m_a, f_a)$ parameter space of ALPs. Another modification of axion or ALP potential is referred as \textit{kinetic misalignment} (\cite{Co:2019jts,Chang:2019tvx,Co:2020xlh}). In this scenario, a higher dimensional explicit PQ breaking potential and a large initial field value for the PQ symmetry breaking field lead to a nonzero initial ALP velocity which in turn, triggers a delayed ALP oscillation bringing modifications in the relic abundance as well as in parameter space of ALPs. A more general treatment of the effect of initial conditions of ALP and its effects on the $(m_a, f_a)$ parameter space can be found in ref. \cite{Eroncel:2022vjg}. Ref. \cite{Chao:2022blc} discusses about the impact of ALP mass modification on its relic abundance, within the context of the type-II seesaw mechanism. 
%\cite{}.

Our proposal differs from the existing works in a sense that it relies on the electroweak symmetry breaking phase (most natural and unavoidable phase) for modifying the ALP potential, without changing its minimum though. The new mass term for ALP that originates at EWSB may provide a dominant or sub-dominant contribution to the effective final mass of the ALP ($m_a$). Depending on the onset of ALP oscillation before or at EWSB, the standard misalignment mechanism gets modified so as to obtain a new parameter space, in $(m_a, f_a)$ plane, carrying significant differences with the standard one which might be interesting from ALP search experiments. It turns out that such a Higgs portal interaction of ALPs allows us to probe for light ALPs. Interactions involving ALP and Standard Model (SM) Higgs have also been exercised in few references \cite{Im:2019iwd,Dev:2019njv,Oikonomou:2023bah}, however in different contexts. Specifically, a higher dimensional operator consisting of axion and Higgs are discussed in \cite{Oikonomou:2023bah} which is responsible for a new minimum of ALP inducing an epoch of kination and generation of gravitational waves. Contrary to our proposal, the ALP there neither plays the role of DM nor obtains a shift in its mass during EWSB.

The outline of the work is as follows. In the next section, we discuss the standard ALP scenario and following this, we move for studying the evolution of ALPs in our modified scenario. The observation and constraints are elaborated in sections \ref{observations} and \ref{constraints} respectively. Finally we conclude in section \ref{conclusion}. 

\section{Standard Misalignment Mechanism and ALP as DM \label{standard}}

In this section, we first elaborate on the production of ALPs in the early Universe via misalignment mechanism followed by standard ALP cosmology and the related parameter space. We presume the existence of ALPs prior 
to the end of primordial inflation as a result of spontaneous breaking of a global $U(1)$ symmetry during inflation. 
After inflation (followed by a reheating era), the ALP field $a ({\vec{x}}, t)$ is expected to be spatially homogeneous, hence described by $a(t)$ only, and gets frozen at an initial value $a_I$ parametrised by the misalignment angle $\theta_I = a_I/f_a$ (with $\dot{\theta}_I = 0$), as long as the Hubble remains larger compared to its mass $m_{a0}$. Note that, the ALP being a Nambu Goldstone Boson, the origin of $m_{a0}$ is related to the breaking of shift symmetry which can possibly be connected to the non-perturbative dynamics, (analogous to PQ axion) and independent of temperature. This can be realised if the ALP couples to a hidden $SU(N)$ sector (the global $U(1)$ symmetry is anomalous under such non-abelian group) which may not be thermalised. 
%However, we continue with $m_{a0} (T)$ in some places of this section for the purpose of generality.

\noindent The related ALP potential can be parametrised by  
\beeq
V_{\rm{0}}= m_{a0}^2 f_a^2\left(1-\textrm{cos} \frac{a}{f_a}  \right), 
\label{V_0}
\eeq
where $f_a$ corresponds to the scale of spontaneous symmetry breaking of the global $U(1)$.
The ALP field $a$ parametrised by $\theta =a/f_a$ follows the classical equation of motion in the background of expanding Universe as given by 
\beeq
\ddot{\theta}+3\mathcal{H}(T)\dot{\theta} - \frac{\nabla^2 \theta}{R^2}+ \frac{1}{f_a^2}\frac{\partial}{\partial \theta} V_{\rm{0}}=0,
\label{ALP_eom-0}
\eeq
and `dot' indicates the derivative with respect to time $t$. 

At very early Universe, after inflation, when $\mathcal{H} \gg m_{a0}$, the $\it{zero ~mode}$ of the axion field remains overdamped and gets stuck at its initial value $\theta_I$ due to Hubble friction. In a radiation dominated era, at some point when 
\beeq
3\mathcal{H}(T^0_{osc}) = m_{a0},
\label{oscillation}
\eeq
the field rolls toward its potential minimum and starts to oscillate. The onset of oscillation can be characterised by the temperature $T^0_{osc}$. Near the minimum of the potential, $V_{\rm{0}} \simeq \frac{1}{2}{m_{a0}}^2 f_a^2\theta^2$ and the equation of motion (below $T \lesssim T^0_{osc}$) reduces to 
\beeq
\ddot{\theta}+3\mathcal{H}(T)\dot{\theta}+ m_{a0}^2 \theta=0, 
\label{ALP_eom}
\eeq
where the ALP field is considered to be a homogeneous one, $\it{i.e.}$ the spatial variation over Hubble volume, 
$\nabla^2 \theta/R^2$ vanishes. Using the expression of the Hubble parameter $\mathcal{H}$ in a radiation-dominated universe, $\mathcal{H}(T)=1.66\sqrt{g_\star(T)}T^2/M_{Pl}$, and the relation in Eq. \ref{oscillation}, we can estimate the conventional oscillation temperature $T^0_{osc}$ for ALP as
\beeq
T^0_{osc}\simeq 1.5\times 10^7 ~\textrm{GeV}~\left( \frac{100}{g_\star(T^0_{osc})}\right)^{1/4}\left(\frac{m_{a0}}{10^{-3}~\textrm{GeV}} \right)^{1/2},
\label{Tosc} 
\eeq
where, $g_\star(T)$ is the number of relativistic degrees of freedom ($\it{d.o.f}$) at a temperature $T$ and the value of Planck mass, $M_{Pl}=1.22\times 10^{19}$ GeV is deployed. 

At the conventional oscillation temperature $T^0_{osc}$, the total energy density of the ALP $\rho_{a} = (\dot{\theta}^2 f^2_a + m_{a0}^2 f_a^2 
\theta^2)/2$ is fully embedded in its potential part only since ALP does not have any initial velocity, $(\dot{\theta}_I=0)$ and is given by 
\beeq
\rho_{a}(T^0_{osc})=\frac{1}{2} m_{a0}^2 f_a^2 \theta_I^2.
\label{rho-a-tosc-0}
\eeq
For $T\lesssim T^0_{osc}$, Eq. \ref{ALP_eom} implies that field would perform fast oscillations with slowly decreasing amplitude where the average energy density $\langle\rho_a\rangle$ scales as $R^{-3}$ and the equation of state $\omega = \langle p_a \rangle / \langle \rho_a \rangle$ turns out to be zero implying that it behaves like non-relativistic matter \cite{Arias:2012az,Marsh:2015xka}. Here, $\langle \rangle$ implies averaging over one complete oscillation. 

Since the ALP number density in a co-moving volume turns out to be conserved \cite{Arias:2012az}, the present day ALP energy density can be estimated as
\bea
\rho_a(T_0) = \rho_{a}(T^0_{osc})\frac{m_a (T_{0})}{m_{a0} (T^0_{osc})}\left( \frac{R_{osc}}{R_0}\right)^3  
\label{ALP_energy-m}
\\
=\frac{1}{2} m_{a0}(T^0_{osc}) m_{a}(T_{0})f_a^2 \theta_I^2\left( \frac{R_{osc}}{R_0}\right)^3,
\label{ALP_energy}
\eea
where $T_0 =2.4\times 10^{-4}$ eV is the present temperature and a change of ALP mass at different (later) temperature, if any, is included in the form of $m_a (T)$. 
However as stated above, unlike QCD Axion, ALP masses are generally considered as temperature independent, 
${\it{i.e.}}~ 
m_{a0} (T^0_{osc})= m_a (T_{0})= m_{a0}$. In this case, 
employing  Eq. \ref{ALP_energy} and considering the adiabatic expansion of the Universe $i.e. ~(R_{osc}/R_0)^3=s(T_0)/s(T^0_{osc})$ where $s$ is the entropy density of the universe given by $s(T)=(2\pi^2/45)g_{\star s}T^3$, the ALP relic density can be expressed as
\beeq
\Omega_a h^2= \frac{h^2}{2\rho_{c,0}}m_{a0}^2 f_a^2 \theta_I^2\left(\frac{g_{\star s}(T_0)}{g_{\star s}(T^0_{osc})} \right)\left( \frac{T_{0}}{T^0_{osc}}\right)^3,
\label{ALP_relic}
\eeq
where $\rho_{c,0}=1.05\times 10^{-5} h^2~\textrm{GeV~cm}^{-3}$ (present critical energy density) and $g_{\star s}(T_0)=3.94$ (number of relativistic $\it{d.o.f}$ at present temperature) \cite{Bauer:2017qwy}. Using the expression of Eq. \ref{Tosc} into 
Eq. \ref{ALP_relic}, the ALP relic density can be estimated as
\beeq
\begin{multlined}
\Omega_a h^2\simeq 0.12\left[\frac{\theta_I}{\mathcal{O}(1)} \right]^2\left[\frac{100}{g_\star(T^0_{osc})}\right]^{\frac{1}{4}}\times\\ \left[\frac{{m}_{a0}}{10^{-9}~\textrm{GeV}}\right]^{\frac{1}{2}} \left[\frac{f_a}{4\times 10^{11}~\textrm{GeV}}\right]^2,
\end{multlined}
\label{con_ALPrelic}
\eeq
from which a correlation between the two parameters $m_{a0}$ and $f_a$ can easily be obtained. However, once the astrophysical and cosmological bounds are imposed (to be discussed later in section \ref{constraints}), the allowed mass range for the ALPs falls below keV-scale \cite{Balazs:2022tjl} only. 

%We provide a plot in Fig. \ref{contour-standard} to show the parameter space in $m^0_a - f_a$ plane for which 
%the correct ALP DM relic can be generated. This correlation is described for different choices of misalignment angle 
%($-\pi \leq \theta_I \leq \pi$) $\theta_I = 1, ~10^{-2}$ represented by green and red lines respectively. 

\section{Higgs portal interaction and ALP \label{observations}}
In this section, we introduce an explicit higher order shift symmetry breaking term in the ALP potential involving the SM Higgs. As a result, a new contribution toward the mass of ALP originates once the electroweak symmetry breaking takes place. The appearance of such a mass term for ALP at an intermediate phase (during its evolution) in addition to $m_{a0}$ (connected to non-perturbative dynamics) not only enables the effective mass of the ALP ($m_a$) and its decay constant ($f_a$) to treat as independent parameters, but also modifies the frequency of ALP oscillation at EWSB as we observe below. 

We first consider the Lagrangian involving a global $U(1)$ symmetry breaking complex scalar field $\Phi = \eta e^{\theta}/{\sqrt{2}}$, as 
\beeq
\mathcal{L} = \frac{1}{2} \left(\partial \eta\right)^2 + \frac{1}{2} f_a^2 \left(\partial\theta\right)^2 - \lambda \left(\eta^2 - f^2_a/2 \right)^2, 
\label{L}
\eeq
\noindent where $\theta = a/f_a$ as parametrised in the earlier section. Once the $U(1)$ global symmetry is spontaneously broken, the potential for the ALP field $a$ is given $V_0$ of Eq. \ref{V_0}. We now introduce additional dimension-6 shift symmetry breaking term involving SM Higgs doublet $H$, as given by 
\beeq
V_1 = \frac{|H|^4}{\Lambda^2}\Phi^2 e^{i\alpha} +h.c.,
\label{V_1}
\eeq
where the phase $\alpha$ in Eq. \ref{V_1} can take values in the range $0\leq\alpha\leq\pi$ \cite{Higaki:2016yqk} and $\Lambda$ acts as a cut-off scale with $\Lambda > f_a$, indicating that the explicit breaking of the global symmetry may take place at some high scale ($\Lambda \lesssim M_{Pl}$) \cite{Draper:2022pvk,Cordova:2022rer}. Based on this known possibility that gravity effects explicitly break a global symmetry \cite{Giddings:1988cx,Coleman:1988tj,Rey:1989mg,Abbott:1989jw,Akhmedov:1992hh} at Planck scale $M_{Pl}$ (or even at a scale much smaller than the Planck one, as recently shown by \cite{Cordova:2022rer} in the context of weak-gravity conjecture \cite{Arkani-Hamed:2006emk}), we discuss the origin of such an operator in Appendix-\ref{Appen:1}. The specific construction mentioned there is also capable of disallowing another dimension-6 explicit $U(1)$ symmetry breaking term $\frac{|H|^2}{\Lambda^2}\Phi^4\exp(i\alpha)+h.c.$ which can be present otherwise. Note that this additional operator can be excluded from a naive consideration too by keeping the order of explicit $U(1)$ breaking (by amount of $U(1)$ charges) minimal. 
While we mostly focus on the explicit breaking operator as in Eq. \ref{V_1} only for the rest of the analysis, we show in Appendix-\ref{Appen:2} that inclusion of the additional dimension-6 operator may lead to an altogether different phenomenology.

The phenomenologically relevant part of the potential for the ALP or the $\theta$ field, after the spontaneous breaking of 
the PQ-like global symmetry, then turns out to be 
\beeq
V_a=m^2_{a0} f_a^2\left(1-\textrm{cos} \frac{a}{f_a}  \right) + \frac{|H|^4}{\Lambda^2}f_a^2 \textrm{cos}\left(\frac{2a}{f_a}+\alpha \right).
\label{V_a}
\eeq
%where $\alpha = \pi$ is invoked.
At a temperature sufficiently higher than the electroweak scale, $T > T_{\rm{EW}} 
\sim$ 150 GeV, the temperature correction to the SM Higgs potential helps the SM Higgs to have a single minimum at origin \cite{Kolb:1990vq}. Hence the Higgs field is expected to settle at origin as a result of which this term does not contribute till the temperature becomes comparable to $T_{\rm{EW}}$ when $H$ gets a vacuum expectation value (vev), $H=(v+h)/\sqrt{2}$ with $v = 246$ GeV. After the EWSB, the ALP receives a new contribution to its mass such that its effective mass $m_a$ satisfies the relation
\beeq
m_a^2 \equiv \left( \frac{d^2 V_a}{da^2}\right) _{{\rm min}}=m_{a0}^2~ {\rm cos}\theta_{\rm min}-\frac{v^4}{\Lambda^2}~ {\rm cos}(2\theta_{\rm min}+\alpha),
\label{m-general}
\eeq
where $\theta_{\rm min}=a_{\rm min}/f_a$ denotes the newly developed minimum of the ALP potential after 
EWSB. The value of $\theta_{\rm min}$ can be estimated as a solution to the equation, 
\beeq
{\rm sin}\theta =\frac{q}{2}~ {\rm sin} (2\theta+\alpha), ~~~ \textrm{with}~q=\frac{v^4/\Lambda^2}{m_{a0}^2},
\label{eq1}
\eeq
obtained by minimising the ALP potential. Clearly, the new potential minimum depends on the choice of $\alpha$ and other parameters ($\Lambda$ and $m_{a0}$). We pursue with $\alpha=\pi$ for the rest of the analysis  
that results into the effective mass of the ALP given by, 
\beeq
m_a^2 = m^2_{a0}+\frac{v^4}{\Lambda^2}.
\label{m-eff}
\eeq
Such a choice of $\alpha=\pi$ is motivated primarily from the fact that the minimum of the ALP potential ($\theta_{\rm min} = 0$) remains unaffected due to the presence of this additional contribution to ALP potential via Eq. \ref{V_1}. However, for conventional QCD axion, such explicitly breaking term should be incorporated very carefully as any alteration of the minimum can spoil the resolution of the strong CP problem, the strong-CP violating angle being heavily constrained by neutron electric dipole moment \cite{Abel:2020pzs} (widely referred as Axion quality problem \cite{Kamionkowski:1992mf,Holman:1992us,Barr:1992qq}). With a general axion like particles, such a solution of strong CP problem is not necessarily be connected to the solution of the strong CP problem. To investigate the impact of 
different $\alpha$ values on $\theta_{\rm min}$ and hence on relic, we incorporate an analysis in the later part of the following sub-section.

Appearance of such a mass term would affect the evolution of the ALP field, in terms of its change in oscillation frequency, immediately after the EWSB. To analyse it further, we divide the study into two cases depending on whether the ALP starts its oscillation at a temperature $T^0_{osc}$ (due to $m_{a0}$) prior to EWSB temperature 
$T_{\rm{EW}}$ or not, as \\
\bea
{\rm{Case}}\left[\rm{A}\right]: &T^0_{osc} > T_{\rm{EW}},\\
{\rm{Case}}\left[\rm{B}\right]: & T^0_{osc} \le T_{\rm{EW}}.
\nonumber
\eea
Note that the demarcation between these two cases is set by the condition 
\beeq
m_{a0} = 3 \mathcal{H} (T_{\rm{EW}}),
\label{m-EW}
\eeq
which translates (assuming a radiation dominated universe) into $m_{a0} > 10^{-13}$ GeV (for case [A]) and $m_{a0} \le 10^{-13}$ GeV (for case [B]) where we use $T_{\rm{EW}} = 150$ GeV.

\subsection{ALP oscillation starts before EWSB}
\begin{figure}[h!]
	\centering
	\includegraphics[scale=0.58]{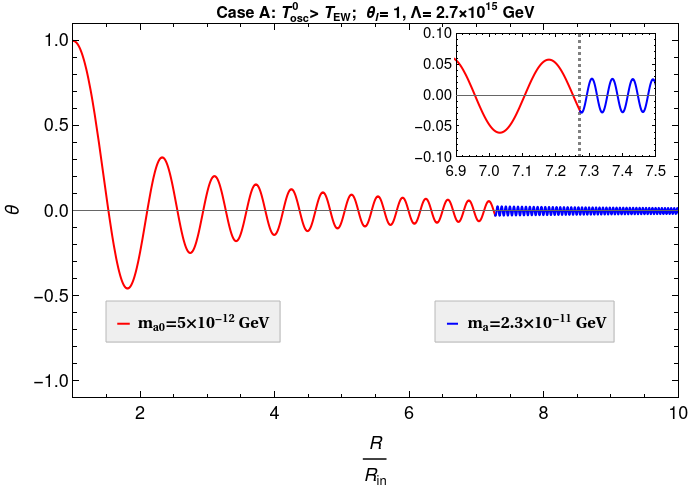}	
	\caption{Evolution of $\theta$ before EWSB (solid red) and after EWSB (solid blue) against ${R}/{R_{\rm in}}$ in case A $(T^0_{osc}>T_{\rm EW})$. In the inset, the evolution of $\theta$ is shown in the vicinity of $T_{\rm EW}$ (the dashed grey vertical gridline of the inset diagram corresponds to $T_{\rm EW}$). }
	\label{caseA-osc}
\end{figure}

Here the ALP is expected to start its oscillation at a temperature $T^0_{osc}$ higher than $T_{\rm{EW}}$, connected to its mass $m_{a0}$. So, the evolution of this ALP for the period $T^0_{osc}$ to $T_{\rm{EW}}$ is guided by Eq. \ref{ALP_eom}. Near the onset of its oscillation at $T=T^0_{osc}$, it carries an energy density same as of Eq. \ref{rho-a-tosc-0}.
At $T=T_{\rm{EW}}$, due to the electroweak symmetry breaking, the higher order Higgs portal interaction provides an additional contribution to its mass as specified in Eq. \ref{m-eff}. As a result, the evolution of the ALP is now governed by the same form of Eq. \ref{ALP_eom}, however replacing $m_{a0}$ by $m_a$ via Eq. \ref{m-eff}, 
$\it{i.e.}$
\beeq
\ddot{\theta}+3\mathcal{H}(T)\dot{\theta}+ m_{a}^2 \theta=0. 
\label{ALP_eom-eff}
\eeq
The shift of ALP mass across the EWSB can be incorporated in estimating the ALP energy density $\rho_a$, crucial in determining the ALP relic density, before and after the electroweak phase transition in line with the discussion 
in the context of Eqs. \ref{ALP_energy-m}-\ref{ALP_energy} as pursued below. The discontinuity of $m_a$ around EWSB can be guided by an appropriate logistic function (as shown in appendix-\ref{Appen:3}) in case one tries to explore evolution of the $\theta$ field as function of scale factor $R$ (normalised by $R_0$), depicted in Fig. \ref{caseA-osc}.

%by imposing the conservation of ALP number density in a co-moving volume as discussed in the context of Eqs. \ref{ALP_energy-m}-\ref{ALP_energy}.}

%{\color{blue} Throughout this work, we have employed the approximation of a sudden change in the ALP mass across EWSB. Such a shift in ALP mass can always be smoothed out by incorporating an appropriate logistic function (see, for instance, Appendix-\ref{Appen:3}), which helps in understanding the evolution of the $\theta$ as function of scale factor $R$ normalised by $R_0$ which is depicted in Fig. \ref{caseA-osc}.}

 Considering the conservation of ALP number density in a co-moving volume across EWSB,
%Now to evaluate the ALP relic density, we proceed to estimate the present energy density of the ALP field $\rho_a$ via Eq. \ref{ALP_energy-m} first. The 
the $\rho_a$ immediately after the EWSB (at $T_{<\rm{EW}}$) can be written, via Eq. \ref{ALP_energy-m}, as
\beeq
\rho_a(T_{<\rm{EW}})=\rho_a(T_{>\rm{EW}})\left( \frac{R_{>}}{R_{<}}\right)^3 \left[ \frac{m_a(T_{<\rm{EW}})}{m_a(T_{>\rm{EW}})}\right].
\label{rho-after-EW}
\eeq
\\
Here $\rho_a(T_{>\rm{EW}})$ is the energy density of the ALP just before the EWSB (at $T_{>\rm{EW}}$) given 
by 
\beeq
\rho_a(T_{>\rm{EW}})=\rho_a(T^0_{osc})\left( \frac{R_{\rm{in}}}{R_{>}}\right)^3, 
\label{rho-before-EW}
\eeq 
where $R_{\rm{in}}$ is the scale factor at the onset of oscillation, $T^0_{osc}$. While the mass of the ALP before EWSB is $m_a({T_{>\rm{EW}}}) = m_{a0}$, $m_a(T_{<\rm{EW}}) = m_a$ results after EWSB as given by Eq. \ref{m-eff}. Here, $R_{>}$ and $R_{<}$ are the scale factors of the universe at temperatures $T_{>\rm{EW}}$ and $T_{<\rm{EW}}$ respectively. 

The energy density of the ALP today (associated with temperature $T_0$) can be written as
\bea
\rho_a(T_0)&=&\rho(T_{<\rm{EW}})\left( \frac{R_{<}}{R_{\rm{To}}}\right)^3\nonumber\\
%&&=\rho_a(T_{osc})\left( \frac{R_{osc}}{R_1}\right)^3\left( \frac{R_{1}}{R_2}\right)^3\left( \frac{R_{2}}{R_0}\right)^3\left( \frac{m_a}{\tilde{m}_a}\right)\nonumber\\
%&&=\rho_a(T_{osc})\left( \frac{R_{osc}}{R_0}\right)^3\left( \frac{m_a}{\tilde{m}_a}\right)\nonumber\\
&&=\rho_a(T^0_{osc}) \frac{m_a}{m_{a0}} \left[\frac{g_{\star s}(T_0)}{g_{\star s}(T^0_{osc})} \right]\left( \frac{T_0}{T^0_{osc}}\right)^3,
\label{caseA_rho0}
\eea
where Eqs. \ref{rho-after-EW}, \ref{rho-before-EW} are employed and $R_{\rm{To}}$ corresponds to today's scale factor.

Using the expressions of $\rho_a(T^0_{osc})$ of Eq. \ref{rho-a-tosc-0}, $T^0_{osc}$ as given by Eq. \ref{Tosc} and plugging them in $\rho_a(T_{0})$ of Eq. \ref{caseA_rho0}, we find the ALP relic density today as
\beeq
\begin{multlined}
\Omega_a h^2\simeq 0.12\left(\frac{\theta_I}{1} \right)^2\left( \frac{100}{g_\star(T^0_{osc})}\right)^{1/4}\times\\
\left(\frac{m_a}{6\times10^{-8}~\textrm{GeV}} \right)  \sqrt{  \frac{10^{-9}~\textrm{GeV}}{m_{a0}}}\left( \frac{f_a}{5\times 10^{10}~\textrm{GeV}}\right) ^2.
\end{multlined}
\label{caseA_ALPrelic}
\eeq
Apart from $m_{a0}$, the $m_a$ dependence relies on the cut-off scale $\Lambda$.  
Unlike the standard dependence of relic on $m_{a0}$ and $f_a$ via Eq. \ref{con_ALPrelic}, here in case-[A], the final relic density also involves the third parameter $\Lambda$, the cut-off scale of the theory which is 
apparent though the involvement of $m_{a}$ in Eq. \ref{caseA_ALPrelic} apart from $m_{a0}$.

\begin{figure}[hb!]
	\centering
	\includegraphics[scale=0.5]{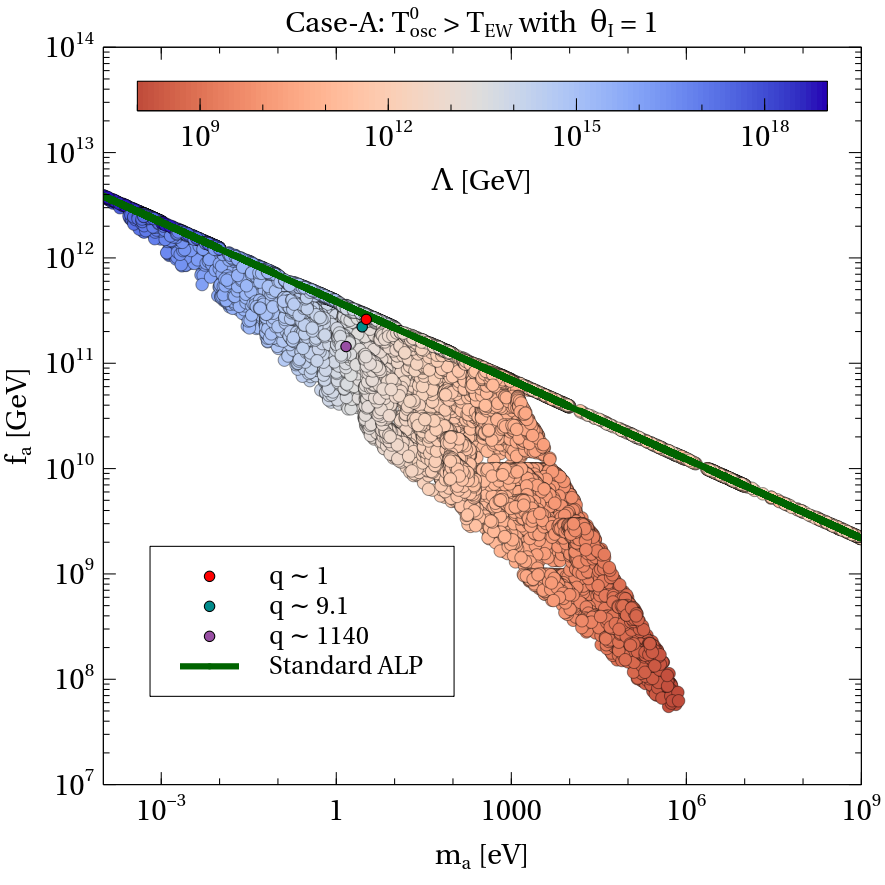}	
	\caption{Relic satisfied parameter space comparison between case A $(T^0_{osc}>T_{\rm EW})$ and standard case in ($m_a-f_a$) plane, while variation of $\Lambda$ is shown in the color bar. The bold dots in red, dark cyan and purple are points taken for studying further dependence of $\theta_{\rm min}$ on $\alpha$.} 
	\label{caseA1}
\end{figure} 

\begin{figure}[hb!]
	\centering
	\includegraphics[scale=0.5]{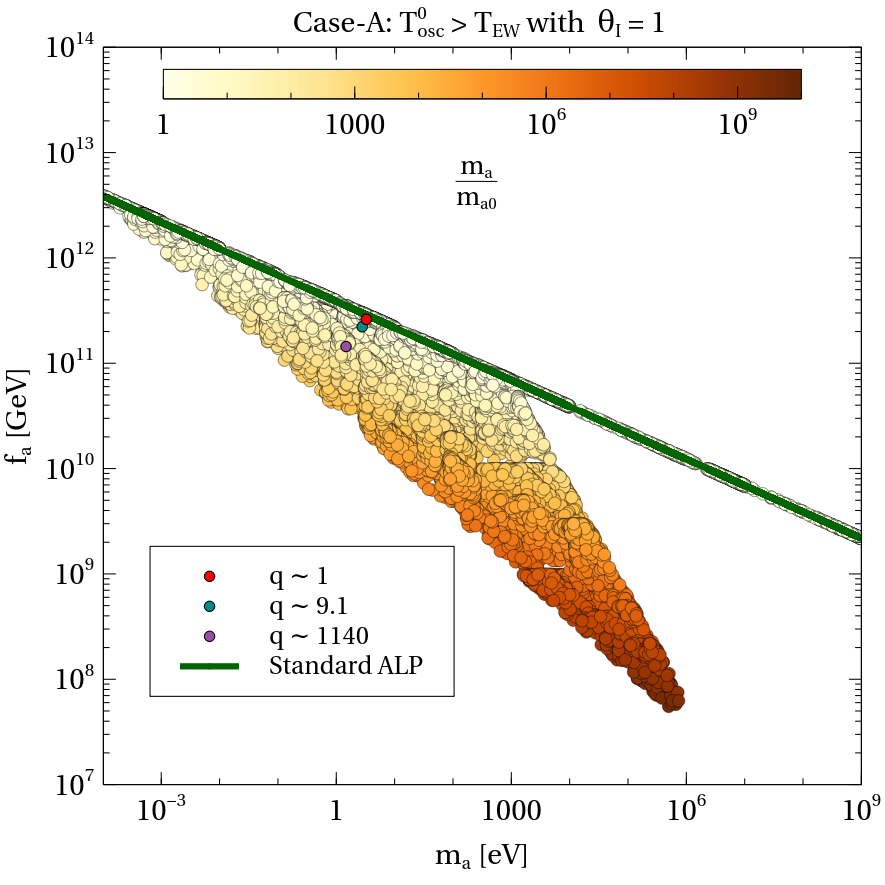}	
	\caption{Relic satisfied parameter space comparison between case A $(T^0_{osc}>T_{\rm EW})$ and standard case in ($m_a-f_a$) plane, while variation of $m_a/m_{a0}$ is shown in the color bar.}
	\label{caseA2}
\end{figure}

The effective mass $m_a$ being the final mass of the ALP which is phenomenologically more relevant than $m_{a0}$, we choose to consider the three parameters as $m_a, f_a$ and $\Lambda$ for our phenomenological analysis. To make the parameters dependence of the relic density explicit, we provide the result of the parameter space scan in $m_a - f_a$ plane shown in Figs. \ref{caseA1} and \ref{caseA2} where the dependence of $\Lambda$ and $m_a/m_{a0}$ are indicated in the respective color maps maintaining $f_a \leq \Lambda \leq 
M_{Pl}$. All the points in $m_{a}$ and $f_a$ plane satisfy the correct relic density $\Omega_a h^2\simeq 0.12$ \cite{Planck:2018vyg}. The gradients of the colors inside the color maps ranging from dark red to blue (for Fig. \ref{caseA1}) and 
yellow to brown (for Fig. \ref{caseA2}) indicate the one to one correspondence between the $\{m_a, f_a\}$ set of values with $\Lambda$ and $m_a/m_{a0}$ respectively for case-[A].
The narrow green line (merged with the borderline of the parameter space of Figs. \ref{caseA1} and \ref{caseA2}) represents the relic-satisfied parameter space for standard scenario with ALP mass, equivalent to $m_a$, from the very beginning. To clarify further, for the green line only, the ALP oscillation begins at some other temperature (say $T_*$) than $T^0_{osc}$ satisfying $3\mathcal{H} (T_*) = m_a(T_*)$. 
In terms of the extended parameter space as obtained in our scenario, this green line acts as the borderline of the parameter space implying that on this line, the change in the ALP mass during EWSB (in presence of the Higgs portal interaction) in very negligible ${\it{i.e.}}~m_a/m_{a0}\simeq 1$, which is clear from Fig. \ref{caseA2}. 
Interestingly, the rest of the extended region allows for a significant gain in ALP mass during EWSB, 
${\it{e.g.}}$ $m_a/m_{a0} \sim \mathcal{O}(10^9)$ for $\{m_a, f_a\} = \{10^{-3}, 5 \times 10^7\}$ GeV, as evident from the top-color-bar of Fig. \ref{caseA2}.

Note that, the presence of the higher dimensional Higgs portal coupling of the ALP allows such broadening of  parameter space (particularly for $f_a$) which would be very significant from experimental perspective clubbed with other constraints which we shall discuss in the next section \ref{constraints}.
%The dependence on the parameter $\Lambda$ (or, $m_a/m_{a0}$) is shown in Fig. \ref{caseA1} (Fig. \ref{caseA2}) in colour bar with dark red towards blue gradient. 
The broadening of the parameter space is an artefact of intermediate change in ALP oscillation frequency as shown in Fig. \ref{caseA-osc}. Considering the ALP to constitute $100\%$ of dark matter energy density in the universe, the upper part of the parameter space in this case (which is merged with the standard case) is excluded by the overabundance of dark matter while the excluded region below this (the left and right border lines of the allowed parameter space) is due to the consideration: $\Lambda>f_a$.
In terms of ALP mass, the lower limit on $m_a$ is kept as $10^{-13}$ GeV here so as to keep $T^0_{osc}$ above $T_{\rm{EW}}$, the higher side of $m_a$ can even be extended beyond the specified value (1 GeV) of the figure. The other constraint, $\Lambda<M_{Pl}$ is only important at the leftmost region of the parameter space  in this case as the minimum contribution from dim-6 operator to ALP mass is $\simeq v^2/M_{Pl} = 5 \times 10^{-15}$ GeV. However, such a correlation involving $m_a$ and the decay constant $f_a$ deserves a further scrutiny from several astrophysical and cosmological bounds which we will discuss in a subsequent section. 

\begin{figure}[ht!]
	\centering
	\includegraphics[scale=0.5]{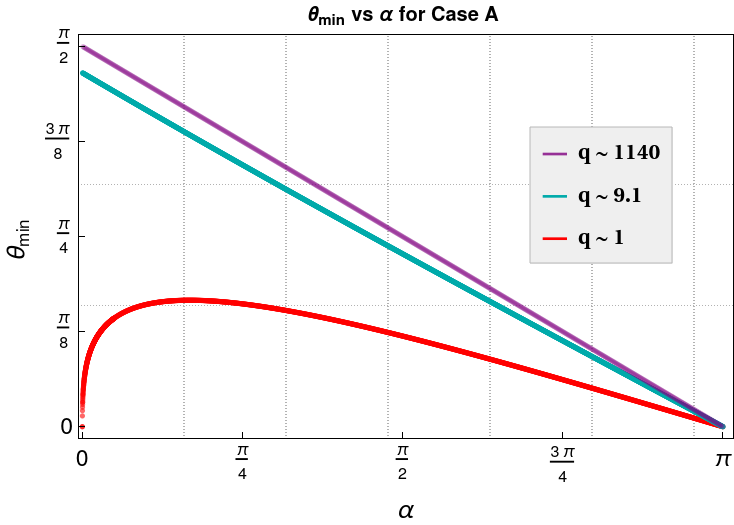}	
	\caption{Variation of ALP potential minimum in terms of $\theta_{\rm min}$ against $\alpha$. The three lines corresponds to three reference points, taken from the parameter spaces in Figs. \ref{caseA1} and \ref{caseA2} having different values of $q$. }
	\label{caseA-thmin-alpha}
\end{figure}

It is perhaps pertinent here to comment on the choice of the phase, $\alpha$. We noted that for the specific choice of $\alpha = \pi$, the $\theta_{\rm min}$ remains at the origin even after the EWSB. However, for an arbitrary $\alpha$, this may not be the case always, indicating a possible impact on the final relic density of ALP. To investigate this, 
we first use Eq. \ref{eq1} to demonstrate the variation of $\theta_{\rm min}$ against $\alpha$ for some choices of  
$q$ values. 
For this purpose we pick up three different sets of parameters $[m_{a0}, f_a, \Lambda]$ (corresponding to three different 
$q$ values) from the parameter space of Fig. \ref{caseA1} (or \ref{caseA2}): 
(a) $[2.31\times 10^{-9},~2.61\times10^{11}, 2.62\times10^{13}] ~{\rm GeV}$, marked in red dark dot having $q~\sim 1$, (b) $[8.74\times 10^{-10},~ 2.22\times10^{11}, 2.29\times10^{13}]~{\rm GeV}$, marked in dark cyan dark dot having $q~\sim 9.1$, and (c)  $[4.35\times 10^{-11}, 1.44\times10^{11}, 4.12\times10^{13}]$~{\rm GeV}, marked in purple dark dot having $q~\sim~1140$. We then employ these three $q$ values to obtain $\theta_{\rm min}$ as function of $\alpha$ for each $q$ as depicted in Fig.  \ref{caseA-thmin-alpha}. 
Firstly we note that with $\alpha=0$, the potential minimum in terms of $\theta_{\rm min}$ depends on the choice of $q$. To be specific, $\theta_{\rm min}$ remains at origin for $q\leq 1$ while it becomes $|\theta_{\rm min}|= {\rm cos}^{-1}(1/q)$ 
for $q > 1$. For non-zero $\alpha$, $\theta_{\rm min}$ picks a distinct value. Hence, a correlation between $\theta_{\rm min}$ and $\alpha$ is noticed in this case as observed in Fig. \ref{caseA-thmin-alpha}. However, for $\alpha=\pi$, $\theta_{\rm min}$ turns out to be zero irrespective of $q$, signifying $\alpha=\pi$ to be an unique choice.

\begin{figure}[ht!]
	\centering
	\includegraphics[scale=0.5]{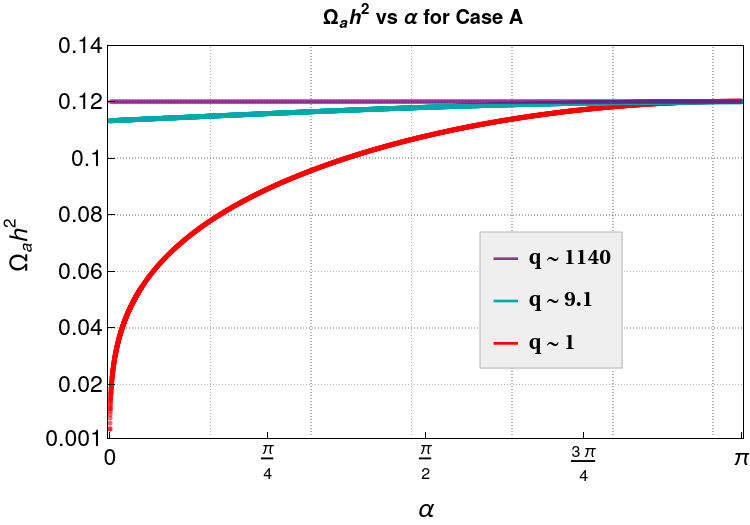}	
	\caption{Variation of the ALP relic density $(\Omega_a h^2)$ vs $\alpha$. The three lines corresponds to three reference points, taken from the parameter spaces in Figs. \ref{caseA1} and \ref{caseA2} having different values of $q$.}
	\label{caseA-relic-alpha}
\end{figure}

Corresponding to Fig. \ref{caseA-thmin-alpha}, a variation in relic density with $\alpha$ is shown in Fig. \ref{caseA-relic-alpha}. It is evident that a maximum relic density results for $\alpha=\pi$. This is simply because the relic density being proportional to the effective ALP mass attains its maximum value for $\alpha = \pi$ as can be seen from Eq. \ref{eq1} (or \ref{m-eff}). Any other choice of $\alpha$ obviously produces a reduced relic density as the resulting mass remains smaller than its maximum value. As indicated in Fig. \ref{caseA-relic-alpha}, the change in relic density is only significant for $q\lesssim \mathcal{O}(10)$. For a sizeable $q$ $({v^4/\Lambda^2} \gg {m_{a0}^2})$, the variation in relic density is almost negligible with $0\leq\alpha\leq\pi$. However, such a decrease in relic density can also be compensated by appropriate scaling of $f_a$ value which is not directly entering in determining the $\theta_{\rm min}$. 

\subsection{ALP oscillation starts after EWSB}

Now we elaborate on the possibility where the ALP is scheduled to start its conventional oscillation (connected to its mass $m_{a0}$ only) after EWSB. As discussed earlier, this can materialise only if $m_{a0} < \mathcal{O}(10^{-13})$ GeV. In this case, even if the global symmetry is spontaneously broken during or before inflation, the ALP field got stuck at the misalignment angle $\theta_I$. The situation may alter with the presence of dim-6 Higgs portal interaction we include in this work leading to two possibilities: (a) the effective mass after EWSB $m_a$ immediately satisfies the condition $m_a(T_{\rm{EW}}) \ge 3\mathcal{H}(T_{\rm{EW}})$, thanks to the Higgs portal contribution toward $m_a$; (b) even with the additional contribution to its mass, $m_a$ satisfies the condition $m_a(T) = 3\mathcal{H}(T)$ with a temperature smaller than $T_{\rm{EW}}$ and hence ALP oscillation starts later.

\begin{figure}[h!]
	\centering
	\includegraphics[scale=0.5]{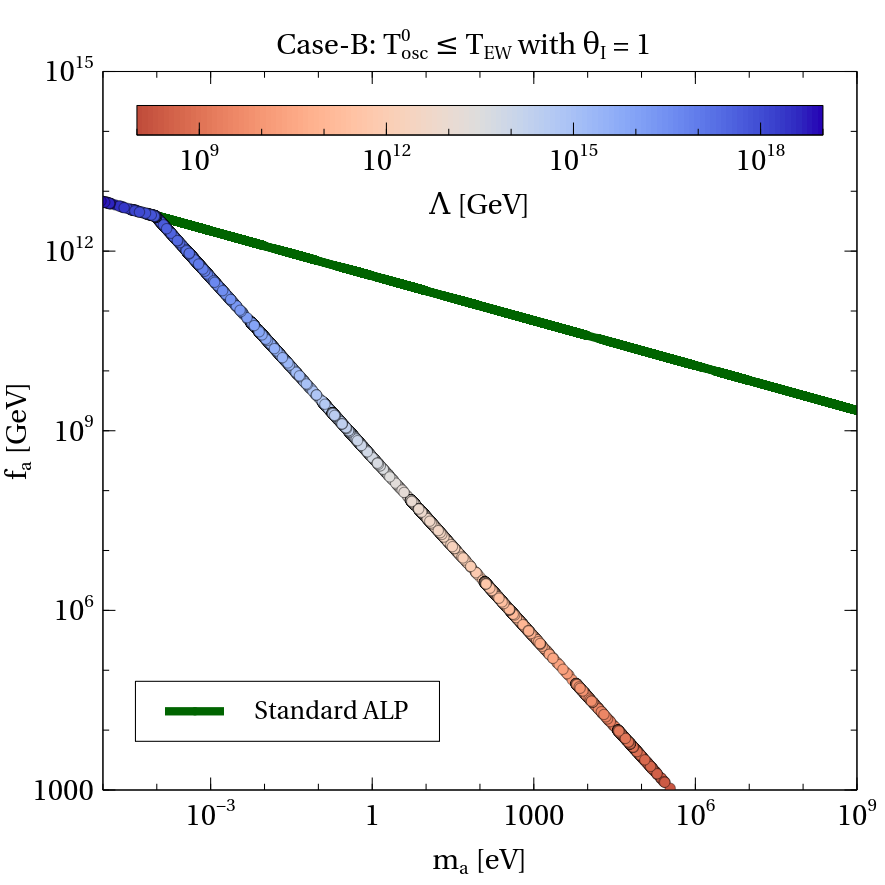}	
	\caption{Relic satisfied parameter space comparison between case B $(T^0_{osc}\leq T_{\rm EW})$ and standard case in ($m_a-f_a$) plane, while variation of $\Lambda$ is shown in the color bar.}
	\label{caseB1}
\end{figure}

\begin{figure}[h!]
	\centering
	\includegraphics[scale=0.5]{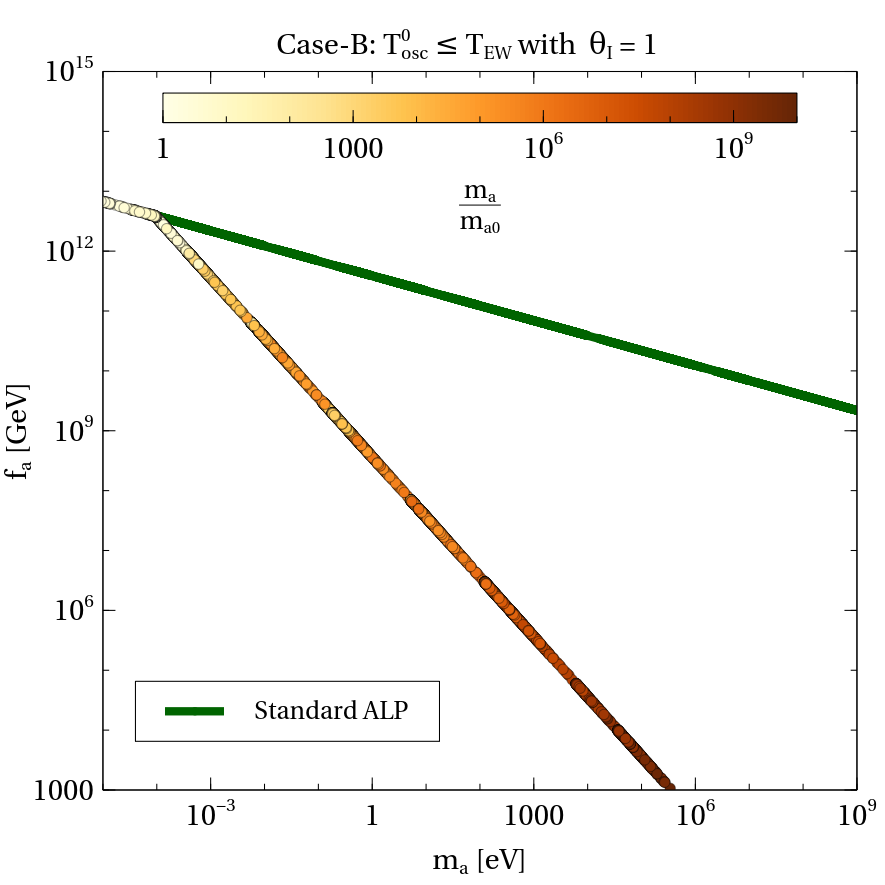}	
	\caption{Relic satisfied parameter space comparison between case B $(T^0_{osc}\leq T_{\rm EW})$ and standard case in ($m_a-f_a$) plane, while variation of $m_a/m_{a0}$ is shown in the color bar.}
	\label{caseB2}
\end{figure}

%\begin{figure}[h!]
%	\centering
%	\includegraphics[scale=0.45]{caseB_osc_new.png}	
%	\caption{Evolution of $\theta$ before EWSB (solid red) and after EWSB (solid blue) against ${R}/{R_{EW}}$ in case B $(T^0_{osc}\leq T_{EW})$. Here $R_{EW}$ is the scale factor at the EWSB and hence the dashed gray gridline at ${R}/{R_{EW}}=1$ represents $T=T_{EW}$.}
%	\label{caseB-osc}
%\end{figure}

We notice that contrary to case [A], there would not be any abrupt change in ALP oscillation here as the oscillation begins already with the effective mass at or below EWSB. The evolution of ALP then proceeds according to the Eq. \ref{ALP_eom-eff}. The rest of the prescription for evaluating the final relic is similar to the standard case discussed in the section \ref{standard}. The $m_a - f_a$ parameter space satisfying the final relic for this case is represented in Figs. \ref{caseB1} and \ref{caseB2} while the corresponding values of $\Lambda$ parameter and the ratio $m_a/m_{a0}$ are shown in top bar, respectively. In the same plot, the standard case ($\it{i.e}$ without Higgs portal coupling) relic satisfied parameter space having ALP mass equivalent of $m_a$ from the beginning is indicated by the green patch for comparison purpose. The relic satisfied parameter space in this case [B] is not broadened (absence of elongated relic satisfied patch as in case [A]) compared to the conventional or standard parameter space as there is no such intermediate change in oscillation frequency. However the standard ALP parameter space for this range of final ALP mass changes its gradient due to a different onset of ALP oscillation era (it starts at $T_* > T_{\rm{EW}}$ related to its mass equivalent of $m_a$). It is found that except for large $\Lambda$ satisfying $\Lambda \sim \mathcal{O}(M_{Pl})$, the Higgs portal operator provides dominant contribution to ALP final mass, ${\it{i.e.}} ~m_a=\sqrt{m_{a0}^2+\frac{v^4}{\Lambda^2}}\simeq \frac{v^2}{\Lambda}$. The minimum contribution to ALP mass obtainable from the dim-6 Higgs portal is of order $v^2/M_{Pl} \simeq 5 \times 10^{-15}$ GeV which sets the boundary of $m_a$ to its lower side. Around this large $\Lambda$, both $m_{a0}$ and Higgs portal contributions are comparable (refer to Fig. \ref{caseB2}) explaining the overlap of the two parameter space lines near $m_{a0} \sim 10^{-13}$ GeV.

%($e.g.$ for two ALPs and Higgs field the coupling reads as $\Lambda_{}$)

\section{Constraints on ALP parameter space \label{constraints}}

In the previous section, we analyse the relic satisfied parameter space of ALP characterized by its final mass 
$m_a$ and the decay constant $f_a$. However, such a parameter space can be further constrained from astrophysical and cosmological limits as well as few laboratory and telescope searches provided one considers 
an ALP-photon coupling \cite{Marsh:2015xka,Jaeckel:2010ni} of the form 
\beeq
\frac{g_{a\gamma\gamma}}{4}F_{\mu\nu}\tilde{F}^{\mu\nu}a,
\eeq
where $F_{\mu\nu}$ and $\tilde{F}^{\mu\nu}$ are the electromagnetic field strength tensor and its dual respectively. The effective coupling $g_{a\gamma\gamma}$ can be written in terms of the ALP decay constant $f_a$ as 
\beeq
g_{a\gamma\gamma}=\frac{\alpha}{2\pi f_a}C_{a\gamma\gamma}.
\eeq
Generally, $C_{a\gamma\gamma}$ is expected to be $\mathcal{O}(1)$ and $\alpha$ is the fine-structure constant.

Such an ALP-photon coupling opens up several windows of observation on which a considerable effort is being devoted now-a-days. These ALPs might get produced within the searing plasma of stars via interactions with photons. Such process may subsequently impact the stellar evolution leading to an overall energy loss of a star while escaping. Therefore, the non-observance of any unwanted energy loss in stars sets bounds on the parameter $g_{a\gamma\gamma}$ \cite{Turner:1989vc,Raffelt:2006cw}. A stringent bound on $g_{a\gamma\gamma}<6.6\times 10^{-11}$ GeV$^{-1}$ emerges from the study of evolution of the horizontal branch (HB) stars \cite{Ayala:2014pea}. Also, the Sun is a likely source of ALPs (solar ALP), which are detectable on Earth in a telescope with a macroscopic magnetic field via reverse Primakoff process, commonly known as the $\it{Helioscope}$ \cite{Sikivie:1983ip}. We have used the latest findings with best sensitivity from the CERN Axion Solar Telescope (CAST) which also puts constraints on $g_{a\gamma\gamma}$ similar to those derived from the study of HB stars as, $g_{a\gamma\gamma}<6.6\times 10^{-11}$ GeV$^{-1}$ for $m_a<0.02$ eV \cite{CAST:2017uph}. The ALP-photon interaction is also constrained by the measurements of solar neutrino flux as $g_{a\gamma\gamma}<7\times 10^{-10}$ GeV$^{-1}$ for $m_a\lesssim \mathcal{O}(\rm{keV})$ \cite{Gondolo:2008dd}. Other important constraints emerge from the cavity experiments such as Rochester-Brookhaven-Florida and University of Florida (RBF and UF) and Axion Dark Matter Experiment (ADMX) which are sensitive for the ALP mass ranges of $4.5-16.3$ $\mu$eV \cite{DePanfilis:1987dk,Wuensch:1989sa,Hagmann:1990tj} and $1.9-3.3~\mu$eV \cite{ADMX:2003rdr} respectively. Telescope searches including Visible Multi-Object Spectrograph (VIMOS) and Multi Unit Spectroscopic Explorer (MUSE) further constrain the ALP mass ranges of $4.5-5.5$ eV and $2.7-5.3$ eV respectively \cite{ParticleDataGroup:2022pth,Regis:2020fhw}. 

\begin{figure}[h!]
	\centering
	\includegraphics[scale=0.5]{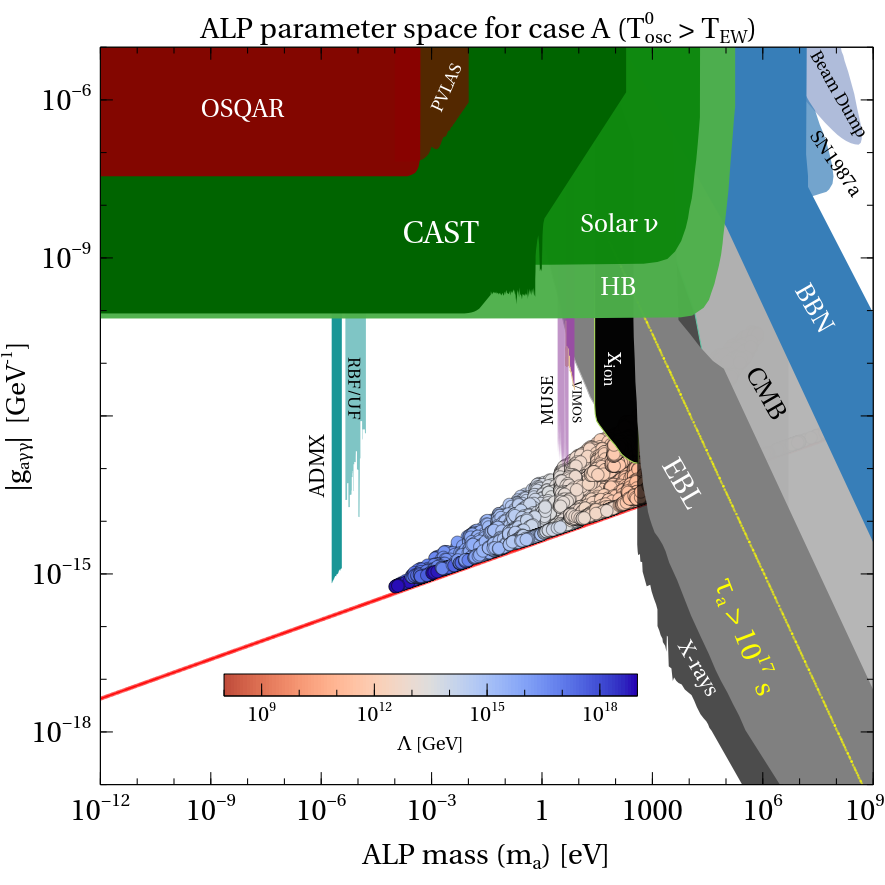}
	\includegraphics[scale=0.5]{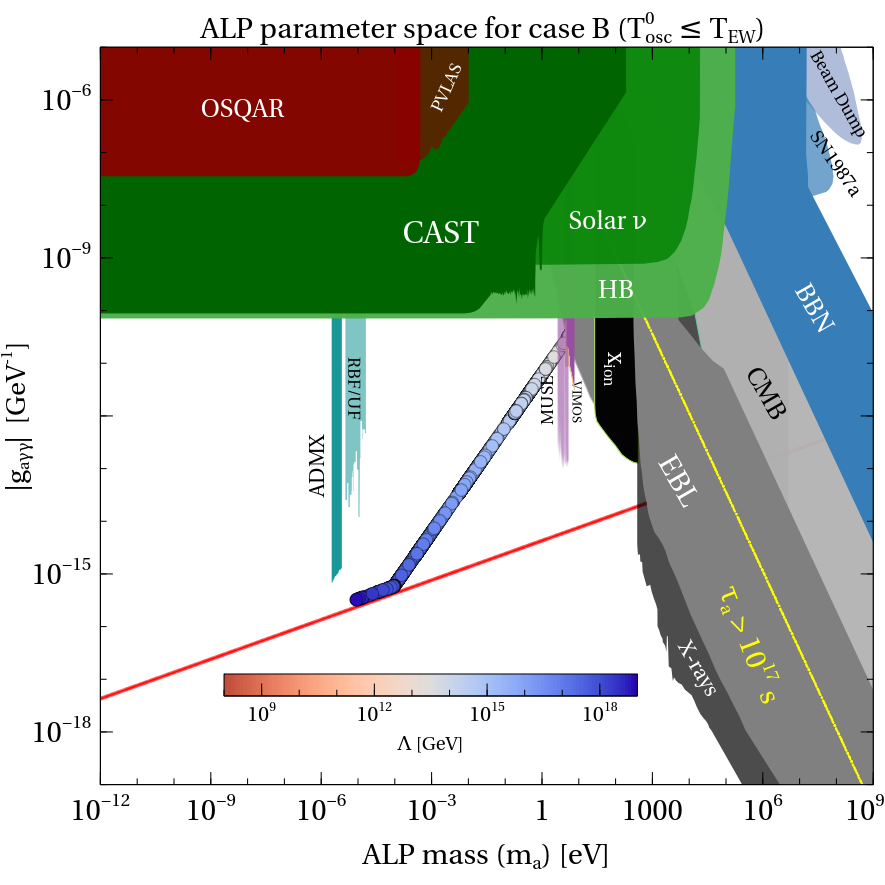}		
	\caption{Excluded regions in ALP parameter space by various constraints along with the relic-satisfied ALP parameter space denoted by dark-red to blue patch for case A (\textit{upper panel}) and blue patch for case B (\textit{lower panel}) in $m_a$-$g_{a\gamma\gamma}$ plane. The solid red line represents the ALP as DM from conventional misalignment mechanism (with $C_{a\gamma \gamma} =1$).}
	\label{ps}
\end{figure}

Searches for ALP are also actively done by various laboratory experiments. One such experimental approach is the light shining through a wall (LSW) experiment \cite{Redondo:2010dp} where a laser beam is expected to be converted into axion or ALPs after being exposed to a high magnetic field. Subsequently, these converted particles pass through an opaque wall and upon re-converting into photons via a second magnetic field behind the wall, they provide indirect evidence for the presence of ALPs. The current best limit by such LSW experiments are given by OSQAR (Optical Search for QED Vacuum Birefringence, Axions, and Photon Regeneration) as $g_{a\gamma\gamma}<3.5\times 10^{-8}$ GeV$^{-1}$ for $m_a<0.3$ meV \cite{Isleif:2022ytq}.

The cosmological constraint of $\Gamma_{a\to\gamma\gamma}^{-1}\geq \tau_U$ (with $\tau_U$ being the age of the Universe and $\Gamma_{a\to\gamma\gamma}=g_{a\gamma\gamma}^2m_a^3/64\pi$ representing the decay width of ALP) serves as a key condition in guaranteeing the stability of the non-thermally produced cosmic ALPs as a viable dark matter over the universe lifetime \cite{Balazs:2022tjl}. If $\Gamma_{a\to\gamma\gamma}^{-1}< \tau_U$, the extra radiations %(encoded in terms of $\Delta N_{eff}$) 
impart additional limits on $g_{a\gamma\gamma}$ extending to very large ALP masses \cite{Masso:1995tw}. Additionally, photons from ALP decays can be seen as peaks on top of the known backgrounds in the galactic x-ray spectra and must not surpass the extragalactic background light (EBL) \cite{Overduin:2004sz}. Also, ALP decaying into photons can lead to the ionization of primordial hydrogen, and the constraint comes from the requirement to prevent this ionization from making a crucial contribution to the optical depth after recombination and hence ensuring the consistency of BBN with observations \cite{Masso:1995tw,Cadamuro:2011fd,Depta:2020wmr}. Other cosmological constraints comes from the excess photons (when decay occurs during opaque universe) include spectral distortions in the CMB spectrum and increase in $T_\gamma$ (photon temperature) relative to $T_\nu$ (neutrino temperature), thus modifying the value of $N_{\rm{eff}}$ inferring from WMAP \cite{Depta:2020wmr}.
	
The bounds regarding all these constraints are shown in the Fig. \ref{ps} which are taken from the updated online repository \verb|AxionLimits| \cite{Ohare2020-gy}. In Fig. \ref{ps}, the yellow line acts as the demarkation line below which the ALP may serve as a viable dark matter ($\Gamma_{a\to\gamma\gamma}^{-1}> \tau_U\simeq10^{17}$ sec). On the other hand, the red line corresponds to the ALP dark matter relic satisfaction contour originating from the standard misalignment mechanism (as discussed in section \ref{standard}) representative of the green line displayed in Figs. \ref{caseA1}, \ref{caseA2}, \ref{caseB1} and \ref{caseB2} with the consideration of $C_{a\gamma\gamma}=1$ (which is followed throughout the analysis). The obtained parameter spaces in our scenario (Figs. \ref{caseA1} and \ref{caseB1}) of cases [A] and [B] are further constrained by these bounds and the residual allowed parameter regions in $m_a - g_{a\gamma \gamma}$ plane (converted from $m_a - f_a$) are demonstrated in top and bottom panels of Fig. \ref{ps} respectively. Note that as seen from Fig. \ref{ps} corresponding to case A, ALP masses 
ranging from $\mathcal{O}$(keV) to $10^{-13}$ GeV (serves as the lower limit for this case as per our previous discussion) are allowed with larger couplings (by several orders of magnitude) compared to the conventional 
picture (red line). Similarly, higher ALP-photon couplings are permitted for case B as well, where the upper limit of allowed ALP mass turns out to be $\mathcal{O}(10)$ eV while the lower limit is set at $m_a\approx 5\times 10^{-15}$ GeV, which is the minimum mass contribution arising from the dimension-6 operator. It is important to note that for standard misalignment of ALP (without any higher dimensional soft symmetry breaking term), the lower limit of ALP mass can be extrapolated upto very small values ($\sim10^{-24}$ eV) \cite{Marsh:2015xka}.

\section{Isocurvature perturbations}

In the present scenario, where the PQ-like symmetry is assumed to be broken during inflation, the ALP field should experience quantum fluctuations having an amplitude denoted by $\delta a\simeq \mathcal{H}_{inf}/2\pi$ $(or,~\delta \theta_I \simeq \mathcal{H}_{inf}/2\pi f_a)$, where $\mathcal{H}_{inf}$ is the Hubble parameter during inflation. These quantum fluctuations give rise to isocurvature perturbation of the cold dark matter \cite{Hamann:2009yf,Beltran:2006sq,Kawasaki:2018qwp} which is constrained from the measurements of the CMB anisotropies\footnote{These ALP fluctuations however do not play any role in the overall density fluctuations of the universe.}. The contribution of ALP to CDM isocurvature perturbation $\mathcal{S}_{CDM}$ can be expressed as
\beeq
\mathcal{S}_{CDM}=\frac{\delta \rho_{CDM}}{\rho_{CDM}}=\frac{\Omega_a}{\Omega_{CDM}}\frac{\delta \rho_a}{\rho_a}.
\label{scdm-0}
\eeq
In our scenario, ALP contributes entirely to relic density of CDM $\it{i.e.}$ $\Omega_{CDM}=\Omega_a$.
The spectrum of CDM isocurvature perturbation in the Fourier space is given by
\beeq
\mathcal{P}_{iso}(k)=\left(|(\mathcal{S}_{CDM})_k| \right)^2,
\label{pcdm}
\eeq
$k$ is the comoving wavenumber, to be evaluated at the pivot scale $k_*$ defined by ${k_*}/a_0=0.05~\textrm{Mpc}^{-1}$. 
The limit imposed by $Planck$ \cite{Planck:2018vyg} on CDM isocurvature perturbation with respect to the adiabatic power, $\mathcal{P}_{adi}(k_*)\approx 2.2\times 10^{-9}$, is expressed as \cite{Planck:2018vyg}
%In this case, the baryon isocurvature is effectively restricted as a form of effective CDM isocurvature. This is because linear-order CMB measurements don't differentiate between baryon and CDM isocurvature modes. The constraint reads as
\beeq
\beta_{iso}=\frac{\mathcal{P}_{iso}(k_*)}{\mathcal{P}_{iso}(k_*)+\mathcal{P}_{adi}(k_*)}<0.038.
\label{iso1}
\eeq
The ALP density perturbation $\mathcal{S}_{CDM}$ of Eq. \ref{scdm-0} can also be recast as
\beeq
\mathcal{S}_{CDM}=2\frac{\delta \theta_I}{\theta_I},
\label{scdm}
\eeq
using ${\delta \rho_a}/{\rho_a}\simeq 2{\delta \theta_I}/{\theta_I}$ which follows from Eq. \ref{rho-a-tosc-0}. Considering the misalignment angle $\theta_I$ to be $\mathcal{O}(1)$ and employing the fluctuation of the misalignment angle during inflation, $\delta \theta_I \simeq \mathcal{H}_{inf}/2\pi f_a$, into Eqs. \ref{scdm} and \ref{pcdm}, we obtain
\beeq
 \mathcal{P}_{iso}=\left(\frac{\mathcal{H}_{inf}}{\pi f_a } \right)^2 \left(\frac{\mathcal{O}(1)}{\theta_I}\right)^2.
\eeq
%Putting $\Omega_ah^2=\Omega_{CDM}h^2=0.12$ and $g_\star(T_{osc})\simeq 100$ into the expression of relic density in Eq. \ref{caseA_ALPrelic} contributed by ALP dark matter, we can obtain the initial value of ALP field expressed as the initial misalignment angle $\theta_i=a_i/f_a$ 
%\beeq
%\theta_i^2\simeq \left( \frac{6\times 10^{-8}~\textrm{GeV}}{m_a}\right) \sqrt{\frac{\tilde{m}_a}{ 10^{-9}~\textrm{GeV}}}\left(\frac{5\times 10^{10}~\textrm{GeV}}{f_a} \right)^2 
%\eeq 
 %Hence,
%\beeq
%\mathcal{P}_{iso}=\left(\frac{H_{inf}}{\pi f_a \theta_i} \right)^2 =\left(\frac{H_{inf}}{1.6\times 10^{11}~\textrm{GeV}} \right)^2 \left( \frac{m_a}{6\times 10^{-8}~\textrm{GeV}}\right) \sqrt{\frac{ 10^{-9}~\textrm{GeV}}{\tilde{m}_a}}
%\eeq
Following the constraint in Eq. \ref{iso1} and considering $\theta_I=1$ as in the previous sections, we obtain an upper bound on the inflationary Hubble scale $\mathcal{H}_{inf}$ as
\beeq
\mathcal{H}_{inf}<2.9\times 10^{-5} f_a.
\label{Hinf-fa}
\eeq
%The ALP mass at the onset of oscillation can be constrained for case [A] and case [B] as we discussed in the following.
Depending on the specific cases in our scenario, we can derive a few more constraints as a consequence of Eq. \ref{Hinf-fa} \cite{Arias:2012az} as we discuss in the following.

For case A ($T^0_{osc}>T_{\rm EW}$), the ALP mass at the onset of oscillation can be constrained by the required condition $3\mathcal{H}_{inf}>3\mathcal{H}(T^0_{osc})=m_{a0}$. Using Eq. \ref{Hinf-fa} and setting $\Omega_ah^2\simeq 0.12$ in Eq. \ref{caseA_ALPrelic}, we obtain the following relation
\beeq
\left(\frac{m_a}{6\times10^{-8}~\textrm{GeV}} \right) \left( \frac{f_a}{5\times 10^{10}~\textrm{GeV}}\right)^{3/2}<6.6\times 10^7,
\label{iso-A}
\eeq
where $g_\star(T^0_{osc})\simeq 100$ is considered.

On the other hand, for case B ($T^0_{osc}\leq T_{\rm EW}$), the constraints can appear in two different ways:
($\it{i}$) when the ALP with effective mass $m_a$ starts to oscillate at $T=T_{\rm EW}$ only, the criteria $3\mathcal{H}_{inf}>3\mathcal{H}(T_{\rm EW})$ along with Eq. \ref{Hinf-fa} leads to 
\beeq
f_a>1.08\times 10^{-9}~\textrm{GeV}.
\label{iso-B1}
\eeq
($\it{ii}$) Secondly, if the oscillation temperature of the ALP with effective mass $m_a$ is itself smaller than $T_{\rm EW}$ (no intermediate change in the ALP mass takes place), we need to utilize the criteria $3\mathcal{H}_{inf}>m_a$. Here, similar to case A, using Eq. \ref{Hinf-fa} and setting $\Omega_ah^2\simeq 0.12$ in Eq. \ref{con_ALPrelic} (with $m_{a0}$ replaced by $m_a$), we obtain 
\bea
f_a&>&1.86\times 10^{8}~\textrm{GeV}.
\label{iso-B2}
\eea

The correlations obtained in Eq. \ref{iso-A} and constraints on $f_a$ as in Eqs. \ref{iso-B1}-\ref{iso-B2} are evidently weaker than the other restrictions shown in the previous section and obeyed by the allowed parameter space in the respective cases. The constraint in Eq. \ref{Hinf-fa} is however significant in the context of gravitational wave detection. As inflation can give rise to the generation of gravitational waves through tensor perturbations, the generation of tensor perturbations during inflation is directly correlated with the Hubble parameter \cite{Kawasaki:2018qwp} as 
\beeq
r=1.6\times 10^{-5}\left(\frac{\mathcal{H}_{inf}}{10^{12}~\textrm{GeV}} \right)^2,
\label{tensor} 
\eeq
which can be translated in our scenario as
\beeq
r<1.34\times 10^{-12} \left(\frac{f_a}{ 10^{13}~\textrm{GeV}} \right)^2 
\eeq
%Hence, for our scenario, the constraint on $H_{inf}$ in Eq. \ref{iso2} translates according to Eq. \ref{tensor} as
%\beeq
%r<3.4\times 10^{-17} \left( \frac{6\times 10^{-8}~\textrm{GeV}}{m_a}\right) \sqrt{\frac{\tilde{m}_a}{ 10^{-9}~\textrm{GeV}}},
%\eeq
which is very small number to be predicted in near future as the current observational constraint on the tensor mode, $r\lesssim 0.1$ , is derived from the Planck measurements of the CMB \cite{Planck:2015sxf}.
\section{conclusion \label{conclusion}}

Axion-like particles are well motivated dark matter candidates which are thought to be produced primarily through the misalignment mechanism in the early universe. In this study, we have explored the impact of electroweak symmetry breaking on the evolution of such ALP DM in presence of an explicit shift symmetry breaking dimension-6 Higgs portal interaction of it. We observe that such an operator may significantly contribute to the ALP mass during the EWSB which further initiates a change in the oscillation frequency, thereby deviating from the standard misalignment mechanism in terms of final outcome. We have shown that depending on the standard ALP oscillation temperature (which is determined by the ALP mass originating from non-perturbative dynamics only), the change in the ALP mass across EWSB gives rise to a significant modifications in relic-density allowed parameter space compared to the standard misalignment mechanism. 

Our findings can be broadly categorised into two: (a) one in which we obtain an extended parameter space (in $m_a - f_a$ plane) compared to the standard misalignment mechanism, applicable when the non-perturbative mass of ALP $m_{a0}$ exceeds $10^{-13}$ GeV and (b) secondly, where we obtain a parameter space with a different slope (in $m_a - f_a$ plane) compared to the standard one which applies when $m_{a0}$ falls below $10^{-13}$ GeV.

Finally, taking into account all the existing constraints from several terrestrial experiments, astrophysical and cosmological bounds on the $m_a-g_{a\gamma\gamma}$ plane (translated from $m_a-f_a$ plane) characterising ALP's interaction with photons, we have identified a significant residue of newly opened up parameter space (from relic satisfaction point of view) in the sub-keV ALP mass regime to be compatible as non-thermal dark matter. These constraints also play a crucial role in restricting the lower limit on the cut-off scale $\Lambda$ (as evident from Fig. \ref{ps}), thereby forbidding the other possibility of ALP production via freeze-in from the decay or annihilations of the Higgs, which requires relatively smaller values of $\Lambda~(\lesssim \mathcal{O}(10^{9})~\textrm{GeV})$. Interestingly, the predicted ALP-photon couplings turn out to be notably larger compared to the case of conventional misalignment, opening up opportunities for exploration through upcoming experiments.

\begin{acknowledgements}
The work of AS is supported by the grants CRG/2021/005080 and MTR/2021/000774 from SERB, Govt. of India. 
\end{acknowledgements}

%%%%%%%%%%%%%%%%%%%%%%%%%%%%%%%%%%%%%%%%%%%%%%%
%Appendix
\appendix
\counterwithin{figure}{section}
\section{Origin of dimension-6 operator}
\label{Appen:1}
Here, we present a possible origin of dimension-6 global $U(1)$ symmetry breaking operator of our kind using the fact that any global symmetry is expected to be explicitly broken by gravity at $M_{Pl}$ \cite{Giddings:1988cx,Coleman:1988tj,Rey:1989mg,Abbott:1989jw,Akhmedov:1992hh}. For this purpose, we consider the following Lagrangian (relevant part only)
\beeq
-\mathcal{L}\supset  \mu \Phi^2 e^{i\alpha} \zeta+ \frac{|H|^4\zeta^\dagger}{M_{Pl}} + h.c..
\label{UV}
\eeq
Here, $\Phi$ is our $U(1)$ symmetry breaking complex scalar field having charge $+1$ under $U(1)$ while $\zeta$ is a heavy (mass $M_{\zeta}$) complex SM singlet scalar field carrying charge -2 such that the first term respects the $U(1)$ symmetry. The dimensionful coupling $\mu$ can naturally be of order $M_{Pl}$. The second term being a dimension-5 operator can be thought of a soft symmetry breaking one resulting due to the explicit breaking of the $U(1)$ by gravity at $M_{ Pl}$. 
Considering the hierarchy of mass scales as: $M_{\zeta}<\mu=M_{ Pl}$, integrating out the heavy field $\zeta$ at energies below $M_{\zeta}$ results into the following dimension-6 term \cite{Bonnefoy:2022vop} 
\beeq
\frac{1}{M_\zeta^2}|H|^4\Phi^2 e^{i\alpha}+h.c..
\eeq
Identifying $M_\zeta$ with the scale $\Lambda$, we obtain the operator considered in our analysis. Note that such a construction would not allow any other dimension-6 operator such as $\frac{|H|^2}{\Lambda^2}\Phi^4 e^{i\alpha}+h.c.$ which could otherwise be present.

\section{Phenemenology in presence of two dimension-6 operators}
\label{Appen:2}
%Even though we focused on the other soft breaking operator (in Eq. \ref{V_1}) for the detailed analysis so far, it is interesting to explore the phenomenology by including both the terms.
	
For completeness purpose, we shall briefly discuss here how the phenomenology changes if we include the other possible dimension-6 operator ($i.e.$ without restricting to the possible origin of such operator as outlined in Appendix-\ref{Appen:1})
\beeq
y\frac{|H|^2}{\Lambda^2}\Phi^4 e^{i\alpha}+h.c.,
\label{V_1alt}
\eeq
where $y$ is a dimensionless coupling. With both the explicit breaking terms present (via Eqs. \ref{V_1} and \ref{V_1alt}) 
and considering $\alpha=\pi$, ALP will acquire a bigger mass (in comparison to Eq. \ref{m-eff}) after EWSB, without altering the potential minimum ($\theta_{\rm min} = 0$), as follows
\beeq
m_a^2= m^2_{a0}+\frac{v^4}{\Lambda^2}+\frac{4y f_a^2  v^2}{\Lambda^2}.
\label{m-eff2}
\eeq
Apart from inducing a sizeable contribution to the final mass of ALP, the new term (Eq. \ref{V_1alt}) also allows a 
dominant production of ALPs through decay (and annihilation) of the Higgs field after EWSB. As a result, a possible 
freeze-in production of ALPs remains viable in addition to the production through misalignment mechanism. Such 
a possibility although prevails with the other dimension-6 operator (we comment on this in the conclusion section), 
here with the additional one via Eq. \ref{V_1alt}, the freeze-in of ALP may proceed with larger coupling strength 
(between two ALPs and the Higgs field) proportional to $v f_a^2/\Lambda^2$, in contrary to the other case (with Eq. \ref{V_1} only), where $f_a$ was replaced by the Higgs vev $v$ ($f_a>>v$). Hence, freeze-in can be significantly enhanced by inclusion of the operator in Eq. \ref{V_1alt}. Additionally, it may lead to the possibility of ALP thermalisation if 
$v f_a^2/\Lambda^2$ happens to be of order $\mathcal{O}(10^{-6})$ or more. To alleviate such possibilities, we find 
that restricting $y$ by $y\lesssim \mathcal{O}(10^{-4})$, ALP can favourably be produced through misalignment mechanism only. 

\begin{figure}[]
	\centering
	\includegraphics[scale=0.5]{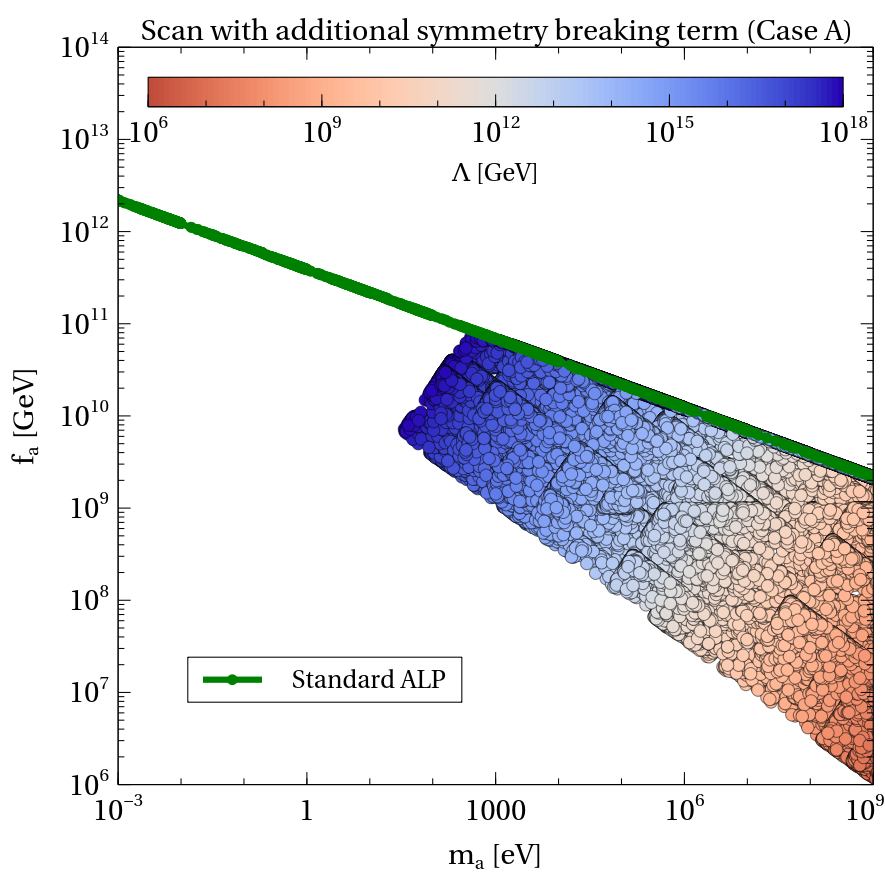}	
	\caption{Relic satisfied parameter space comparison between case A $(T^0_{osc}\leq T_{\rm EW})$ and standard case in ($m_a-f_a$) plane in presence of two dimension-6 $U(1)$ breaking terms. Here, the variation of $\Lambda$ is shown in the color bar and $y=10^{-4}$ is employed.}
	\label{spl-case}
\end{figure} 

Therefore, considering the presence of both the dimension-6 operators with $y\lesssim \mathcal{O}(10^{-4})$, we showcase the relic allowed parameter space in Fig. \ref{spl-case}. Here $m_{a0} > \mathcal{O}(10^{-13})$ GeV is taken so that the conventional ALP oscillation starts prior to EWSB. As the operator of Eq. \ref{V_1alt} adds a substantial contribution to ALP mass via Eq. \ref{m-eff2}, a lower value of $f_a$ is required to achieve the correct relic density as evident from Eq. \ref{caseA_ALPrelic}, which is also observed in Fig. \ref{spl-case}. We notice that with the choice of $y=10^{-4}$, the relic allowed parameter space is shifted beyond $m_a\gtrsim \mathcal{O}(100)$ eV, where the leftmost boundary is dictated by the choice of $\Lambda<M_{Pl}$. Hence it turns out that after imposing all the relevant cosmological and astrophysical bounds (as outlined in section \ref{constraints}), only a tiny fraction of the parameter space remains viable to achieve the correct relic density through misalignment mechanism in this case.

\section{Transition of ALP mass across EWSB}
\label{Appen:3}

\begin{figure}[b!]
	\centering
	\includegraphics[scale=0.5]{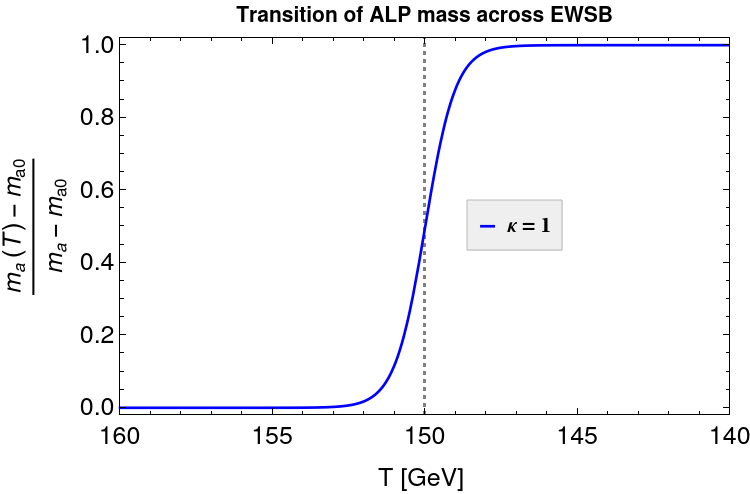}	
	\caption{Transition of ALP mass across $T_{\rm EW}=150$ GeV with $\kappa=1$. The dashed gray gridline corresponds to $T=T_{\rm EW}$.}
	\label{mass-transition}
\end{figure}

In connecting the evolution of the ALP field $\theta$ across $T_{\rm{EW}}$, an apparent discontinuity is felt due to the sudden change in the mass of ALP: from $m_{a0}$ at $T < T_{\rm EW}$ to $m_a$ at $T = T_{\rm{EW}}$ (representative of a step function at $T_{\rm EW}$). To retain the continuity of the ALP field while passing through the EWPT, we propose to make the transition of $m_{a0}$ to $m_{a}$ a smooth one by introducing logistic function for the mass $m_a (T)$ \cite{DiBari:2020bvn} defined as 
\beeq
m_a (T) = m_{a0} + \frac{m_a(T) - m_{a0}}{1 + e^{-2\kappa \left(T_{\rm{EW}} - T\right)}}, 
\label{maa-func}
\eeq
across $T = T_{\rm{EW}}$. Here, $\kappa$ is a parameter which controls the abruptness involved in the transition (a large $\kappa$ indicates a more sharper transition). Since the EWPT happens within a finite range of temperature $\Delta T$, such an approximation is justified provided the the change of ALP mass extends over that period $\Delta T$. A schematic presentation of such an approximation is exhibited in Fig. \ref{mass-transition} with $\kappa=1$ is employed. This is for the illustration purpose in case one tries to track the evolution of the $\theta$ field via Eq. \ref{ALP_eom-eff}. While evaluating the energy density of the ALP field, we use the sudden approximation in line with Eq. \ref{rho-after-EW}.

\bibliography{ref.bib}

\end{document}